\documentclass{camera}
\usepackage{epsfig,amsmath}

\def\ep{\epsilon}

\newcommand\as{\alpha_{\mathrm{S}}}
\def\jet{\rm jet}
\def\ltap{\raisebox{-.4ex}{\rlap{$\,\sim\,$}} \raisebox{.4ex}{$\,<\,$}}

\begin{document}
\renewcommand{\thefootnote}{\fnsymbol{footnote}}
%
\title{HIGGS BOSON PRODUCTION\\ AT TEVATRON RUN II AND LHC
$^*$
}

%
\author{M. Grazzini}


\vspace*{-1cm}

\organization{Dipartimento di Fisica, Universit\`a di Firenze\\ and\\ INFN, Sezione di Firenze, Largo Fermi 2, I-50125 Firenze, Italy}

\maketitle

\vspace*{-1cm}

\abstract{The main production channels of the Higgs boson at hadron colliders are
briefly
reviewed and recent developments in the calculation of QCD effects are discussed.} 

%
\footnotetext[1]{Invited talk given at the XIII italian meeting on high energy physics ``LEPTRE'', Rome April 18-20 2001, to appear in the proceedings.}

\renewcommand{\thefootnote}{\arabic{footnote}}

\vskip 1cm

The Higgs boson is an essential ingredient in the Standard Model, but it has not yet been observed.
After the end of LEP program, the Higgs search will be carried out at hadron colliders. In this talk I will discuss what are the main channels
in which the Higgs can be produced and what is the status of the calculation of QCD corrections.

$\!\!\!\!\!\bullet~gg\to H$\\
The gluon-gluon fusion through a heavy quark loop is the
dominant production
mechanism
at hadron colliders.
At the Tevatron Run II \cite{teva} it leads to about $65\%$ of the total cross section in the range $100$--$200$ GeV.
At the LHC \cite{spira}
$gg$ fusion
dominates over the other production
channels for a light Higgs and at $M_H\sim 1$ TeV still provides about $50\%$ of the total production rate.

The NLO QCD corrections to this process have been computed \cite{ggnlo} and they give
a large effect
increasing the cross section for the production of a light Higgs
of $\sim 100\%$ ($\sim 90\%$)
at the Tevatron Run II (LHC).
Unfortunately at the Tevatron, at least for $M_H\ltap 135$ GeV,
this channel is
swamped
by the huge QCD background, and the production rate is
too small to observe the rare $H\to\gamma\gamma$ decay \cite{teva}. 

$\!\!\!\!\!\bullet~q{\bar q}\to q{\bar q} V^*V^*\to q{\bar q}H$\\
In this channel the Higgs is produced through the fusion of two vector bosons. 
The QCD corrections have been computed within the structure function approach \cite{Han:1992hr} and they increase the cross section of about $10\%$ both at the Tevatron \cite{teva} and at the LHC \cite{spira}.
This channel does not seem to be promising at the Tevatron Run II and at the LHC becomes competitive with $gg\to H$ for $M_H\sim 1$ TeV.

$\!\!\!\!\!\bullet~q{\bar q}\to V\to VH$\\
This channel is the most promising at the Tevatron
for $M_H\ltap 135$ GeV, where the $b{\bar b}$ decay is dominant. This is due to the possibility to trigger on the leptonic decay of the vector boson. The QCD corrections are the same as for Drell--Yan \cite{dy} and increase the cross section of $\sim 30\%$ at the Tevatron Run II \cite{teva} and of $25$--$40 \%$ at the LHC \cite{spira}.

$\!\!\!\!\!\bullet~q{\bar q},gg\to H Q{\bar Q}$
\vskip .2cm
$~i)$  $Q=t$. QCD corrections are known in the limit $M_H\ll m_{top}$ \cite{Dawson:1998im}. In this limit the cross section factorizes in the
convolution
of the $t{\bar t}$ cross section with a splitting function $t\to t H$. However this result is not expected to be quantitatively reliable for realistic Higgs masses.

$~ii)$  $Q=b$. QCD corrections have been computed at NLO both in $\as$ and in $1/\log M_H/m_b$ \cite{Dicus:1999hs} and their effect is separately large but they tend to compensate each other to give a total small effect.
However, in order to isolate the signal, one should observe
one $b$ or both at large transverse momentum.
This process is known only at LO and it
would be very important to have the NLO QCD corrections in order to perform realistic simulations.

In summary, both these channels give small production rates. The $Hb{\bar b}$ channel can be more important in new physics scenarios where the $Hb{\bar b}$ coupling is enhanced. An example of such a scenario is the MSSM with large $\tan\beta$. 

At the LHC the $t\bar t H$ channel can complement the $WH$ one (with $W$
decaying leptonically) in the search
for a light SM Higgs boson, $M_H\ltap 130$ GeV, by triggering on the
leptonic decay of one of the top, while reconstructing the other in the
hadronic decay mode.


In the following I will concentrate on the $gg\to H$ channel.
Since the NLO QCD corrections are quite large, the calculation at NNLO would be very important.
However being a three-loop calculation, it is certainly very difficult.
The large $m_{top}$ approximation
allows to replace the heavy-quark loop through which the Higgs is produced in an effective vertex,
and thus to reduce by one the number of loops. The approximation has been shown
to work at NLO within $5 \%$ for $M_H\ltap 2m_{top}$ \cite{spira}.

In Ref.\cite{Catani:2001ic,Harlander:2001is} the soft and virtual NNLO
corrections to Higgs boson production in the large $m_{top}$ approximation were presented.
The calculation of Ref.\cite{Catani:2001ic} was done by combining the recent
results \cite{Harlander:2000mg} for the two-loop amplitude $gg\to H$
in the large $m_{top}$ limit with the soft factorization formulae for tree-level
\cite{softtree} and one-loop \cite{softloop}
amplitudes.
The independent calculation of Ref.\cite{Harlander:2001is} was performed
with a different method
and the analytical results fully agree.
From the theoretical side the result is important since it provides
a first check of the cancellation of the IR poles from $1/\ep^4$ to $1/\ep$ between real and
virtual contributions.

In Ref.\cite{Catani:2001ic} the hadronic cross section was evaluated consistently at NNLO using the recent MRST2000 set that includes (approximated) NNLO densities \cite{Martin:2000gq}.
Our results provide two estimates of the NNLO cross section: the soft-virtual (SV) and soft-virtual-collinear (SVC) approximation \cite{Catani:2001ic}.
In the SV approximation only the contributions of soft and virtual origin are taken into account. 
This approximation certainly gives the dominant contribution when $\tau=M_H^2/S\to 1$. However, even for small $\tau$ the SV approximation works very well.
In fact the parton distributions are strongly suppressed at large $x$ and
thus the partonic cross section is almost always evaluated close to threshold.
Nevertheless we find that subleading
contributions of purely {\em collinear} \cite{Kramer:1998iq}
origin are numerically important. 
Thus the SVC approximation is defined including the leading logarithmic correction from the collinear region in the $gg$ channel.

The results show a nice reduction of scale dependence at NNLO
(from $\pm 20 \%$ at NLO to $\pm 10 \%$ at NNLO-SV).
At the LHC the NNLO corrections enhance the cross section from $10$ to $25\%$
for a light Higgs with respect to NLO ($K\sim 2.2$--$2.4$).
At the Tevatron Run II ($M_H=150$ GeV) the NNLO effect is more sizable, increasing the cross
section of about $50\%$ with respect to NLO ($K\sim 3$). This large
effect is expected since we are closer to threshold. 
The large $K$-factor at the Tevatron Run II could help for the detection of a Higgs boson in the mass
range $140$--$180$ GeV. 

At the LHC in the mass range $120$--$140$ GeV the $b{\bar b}$ decay mode is overwhelmed by the QCD background and one is forced to look at the $\gamma\gamma$ decay mode, with small branching ratio (${\cal O}(10^{-3})$).
The $pp\to H+\jet$ channel was proposed with the aim of improving the
situation in the $\gamma\gamma$ decay mode \cite{Abdullin:1998er}.
In fact this channel offers several advantages: the photons are more
energetic than in the inclusive channel and the reconstruction of
the jet should allow a more precise determination of the interaction vertex. 
Moreover the presence of the jet allows a better suppression of the
background. This advantages should be able to 
compensate the loss in the production rate. 

In Ref.\cite{deFlorian:1999zd} the NLO QCD corrections
to this process were computed in the large $m_{top}$ limit.
This approximation is expected to work provided both the transverse momentum $p_T$
and the Higgs mass $M_H$ are smaller than $m_{top}$.
The tree level and one-loop amplitudes needed for this calculation
were computed in Refs.\cite{amplitudes}. They were implemented in a Monte Carlo
program using the subtraction method
to handle and cancel infrared singularities \cite{sub}.
This program allows to study any infrared safe quantity for this process at NLO.
The results show that the scale dependence is reduced from $\pm 35\%$ to $\pm 20\%$ going at NLO, and that the $K$-factor is roughly constant with respect
to the kinematics and about $1.6$ \cite{deFlorian:1999zd}.

Let us finally consider the inclusive $p_T$-spectrum of the Higgs boson.
The calculation performed in  Ref.\cite{deFlorian:1999zd} is reliable
only in the region $p_T^2\sim M_H^2$. When $p_T^2\ll M_H^2$ large logarithmic
corrections of the form $\as^2\log^{4}M_H^2/p_T^2$ appear that have
to be resummed to all orders.
The resummation is usually performed in the impact parameter $b$-space \cite{resum} and the
large logarithmic corrections are exponentiated in the Sudakov form factor\footnote{For more details and recent theoretical progress see Ref.\cite{uni}.}:
\begin{equation}
S(M_H,b)=\exp \left\{ -\int_{b_0^2/b^2}^{M_H^2} \frac{dq^2}{q^2} 
\left[ A(\as(q^2)) \;\ln \frac{M_H^2}{q^2} + B(\as(q^2)) \right] \right\}\, .
\end{equation}
The coefficients $A^{(1)}$, $B^{(1)}$ and $A^{(2)}$ that control the
resummation at NLL level were computed in Ref.\cite{Catani:1988vd}.
The most recent phenomenological analysis is performed in Ref.\cite{lhcproc}.
The NLL resummed result, valid in the low $p_T$ region is matched with
the NLO calculation of Ref.\cite{deFlorian:1999zd}.

Recently, the NNLL coefficient $B^{(2)}$ was computed \cite{deFlorian:2000pr}.
This result will certainly allow to improve the matching between resummed
and fixed-order calculations\footnote{A preliminary estimate shows that
the numerical effect of $B^{(2)}$ should be quite large \cite{bal}.}.
Moreover the knowledge of the coefficient $B^{(2)}$, together with the recent numerical
estimate of the coefficient  $A^{(3)}$ \cite{Vogt:2001ci} will allow a (partial) extension
of the accuracy of this calculation to NNLL.

\noindent {\bf Acknowledgments.}
I wish to thank Stefano Moretti for
helpful discussions.

\end{document}